\documentclass[aps,prb,reprint,showpacs,superscriptaddress]{revtex4-1}
\usepackage{graphicx}
\usepackage{ulem}
\usepackage{dcolumn}
\usepackage{bm}
\usepackage{amssymb}
\usepackage{color}
\usepackage{amsmath}
\usepackage{amsfonts}

\newcommand{\be}{\begin{equation}}
\newcommand{\ee}{\end{equation}}
\newcommand{\bea}{\begin{eqnarray}}
\newcommand{\eea}{\end{eqnarray}}
\newcommand{\bse}{\begin{subequations}}
\newcommand{\ese}{\end{subequations}}

\newcommand{\tK}{\tilde K}
\renewcommand{\l}{\ell}
\newcommand{\comment}[1]{}

\begin{document}

\title{Singular response to a dopant of an evaporating crystal surface}

\author{Vladislav Popkov}
\email{popkov@fi.infn.it}
\affiliation{Dipartimento di Fisica e
Astronomia, Universit\`a di Firenze, via G. Sansone 1, 50019 Sesto
Fiorentino, Italy}
\affiliation{Max Planck Institute for Complex
Systems, Nothnitzer Strasse 38, 01187 Dresden, Germany}
\affiliation{Istituto dei Sistemi Complessi, Consiglio Nazionale
delle Ricerche, Via Madonna del Piano 10, 50019 Sesto Fiorentino,
Italy}
\author{Paolo Politi}
\email{paolo.politi@isc.cnr.it}
\affiliation{Istituto dei Sistemi Complessi, Consiglio Nazionale
delle Ricerche, Via Madonna del Piano 10, 50019 Sesto Fiorentino,
Italy}
\date{\today}
\pacs{68.55.-a, 05.70.Ln, 81.16.Rf}

\begin{abstract}
Moving crystal surfaces can undergo step-bunching instabilities, when
subject to an electric current. We show analytically that an infinitesimal
quantity of a dopant may invert the stability, whatever the sign of the
current. Our study is relevant for experimental results [S. S. Kosolobov
\textit{et al.}, JETP Lett. \textbf{81}, 117 (2005)] on an evaporating
Si(111) surface, {which show a singular response to Au doping},
whose density distribution is related to inhomogeneous Si diffusion.
\end{abstract}

\maketitle



\renewcommand{\l}{\ell}


\textit{Introduction}.---Crystal surfaces may undergo dynamical
instabilities while growing or evaporating~\cite{reviewPP,reviewCM}. An
important class of instabilities is that determined by an asymmetric current
of adatoms diffusing on terraces. This asymmetry may be intrinsic, e.g.
determined by asymmetric attachment to steps~\cite{asy_steps} or by surface
reconstruction~\cite{Zhao_Rapid}, or it may be extrinsic, e.g. due to
impurities~\cite{impurities} or to an electric current~\cite%
{electromigration}. Extrinsic instabilities may have the advantage to be
tunable and therefore to be suitably switched on and off. In this Letter we
consider an example of extrinsic instability, displaying a phenomenon which
is of double interest. An interest for fundamental nonequilibrium physics,
because we propose a model showing a singular response to doping. But also
an interest for applications, because we suggest that recent experimental
results~\cite{Latyshev} on evaporating Si(111) surfaces under an electric
current, {showing a singular response when}
exposed to a variable quantity of Au, can be understood at the
light of our model.

Vicinal surfaces are obtained cutting a crystal along an orientation which
is close to a high symmetry one, resulting in a morphology similar to a
flight of steps. During evaporation, atoms detach from steps and diffuse on
terraces until possible desorption. The application of an electric current
induces an electromigration force, whose direction may depend on temperature
$T$: this is the case for the widely studied case of Si(111), where a
sequence of stable and unstable regions are found when varying $T$~\cite%
{sequence}. Here we are interested to analyze how stability is affected by
doping. We assume (and later we discuss) that doping induces inhomogeneity
in the diffusion process, both in its symmetric part and in the drift term,
the latter modifying the electromigration force. In particular, we show that
a divergent drift in a region of vanishing size may induce a change of
stability, even if the sign of the drift does \textit{not} change. This
result might not appear surprising as it is, because step decoration~\cite%
{decoration}, changing attachment kinetic coefficients may have a similar
effect. However, step decoration misses two important features: first, in
our model and in the Si(111) experimental system, doping induces a change of
stability for both signs of the current, meaning that doping cannot be
reduced to a fixed stabilizing or destabilizing effect; second, an
increasing quantity of dopant may induce further stability/instability
transitions. In this Letter we focus on the very first transition, occurring
at a vanishing critical density of the dopant. However, we also argue how a
larger quantity of dopant may further change the stability of the surface.

\begin{figure}[ptb]
\includegraphics[width=\columnwidth,clip=yes]{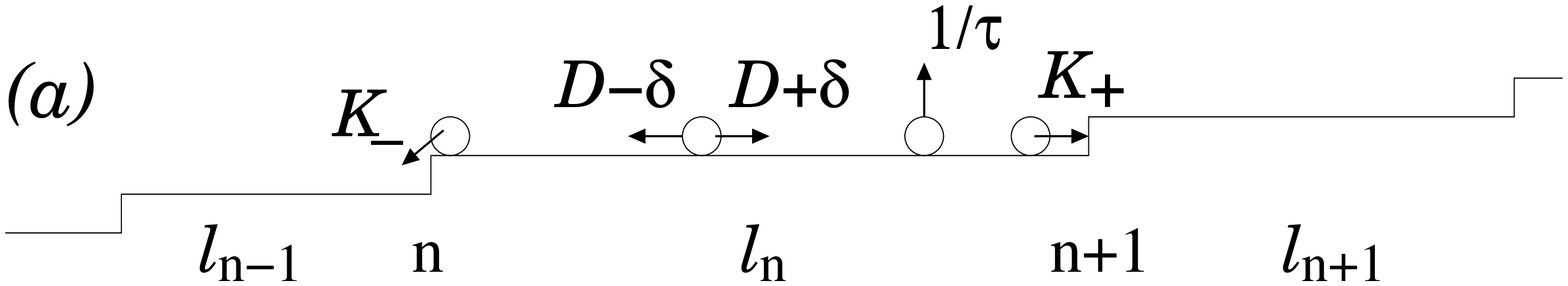} %
\includegraphics[width=\columnwidth,clip=yes]{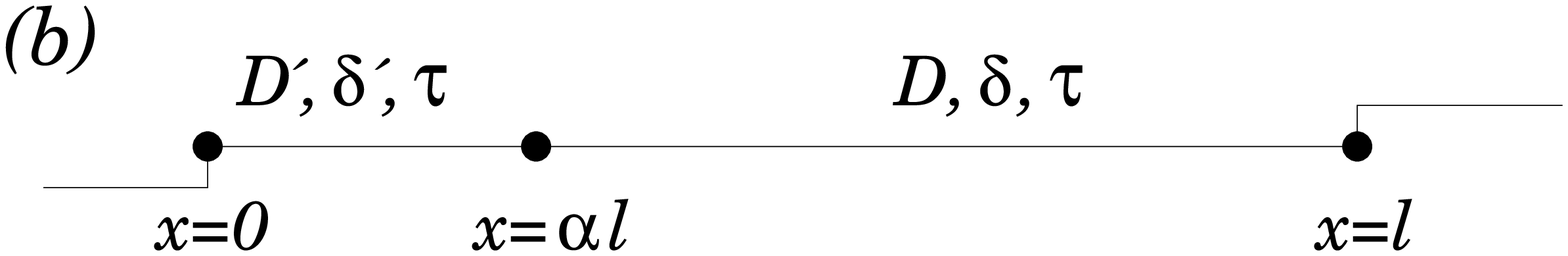}
\caption{(a) Profile of a one dimensional surface and (b) profile of a
terrace, relevant for the inhomogeneous model. All symbols are explained in
the text.}
\label{sup_vicinale}
\end{figure}

\textit{The model}.---The simplest model of adatom diffusion in the presence
of evaporation and drift (see Fig.~\ref{sup_vicinale}a) requires to solve
the following stationary diffusion equation for the adatom density $p(x)$,
\begin{equation}
Dp^{\prime \prime }(x)-2{a_0^{-1}}\delta p^{\prime }(x)-\gamma ^{2}p(x)=0,
\label{eq_diffusione}
\end{equation}%
{where  $a_0$ is the lattice constant; $D$ is the diffusion
constant; $\delta/a_0= q_{\hbox{\tiny eff}}E D/(k_B T)$ is the
drift, with $q_{\hbox{\tiny eff}}$ being the effective charge of
adatoms and $q_{\hbox{\tiny eff}}E$ being the electromigration
force; $\gamma ^{2}=1/\tau $ is the evaporation rate.} As
illustrated in Fig.~\ref{sup_vicinale}b, we adopt a simple model
for inhomogeneity: each terrace is separated in two regions with
different values for $D$ and $\delta $. Boundary conditions are
expressed in terms of the surface current
\begin{equation}
J(x)=-Dp^{\prime }(x)+2{a_0^{-1}}\delta p(x),
\label{surface_current}
\end{equation}%
 which at steps is
proportional to the supersaturation,
\begin{equation}
J_{\hbox{\tiny step}}=\pm K(p_{\hbox{\tiny step}}-p_{\hbox{\tiny
eq}}). \label{BCusual}
\end{equation}%
Here the plus (minus) sign applies to the ascending (descending)
step. We expressly chose the same kinetic coefficient $K$ for
ascending ($K_{+}$) and descending ($K_{-}$) step, to avoid
stabilizing or destabilizing effects due to their asymmetry,
$K_{+}\neq K_{-}$~\cite{reviewPP,reviewCM}.

The general solution of Eq.~(\ref{eq_diffusione}) for the piece of
terrace $x \in [\alpha \l,\l]$ is given by
\begin{equation}
p(x)=Ae^{\lambda _{1}x}+Be^{\lambda _{2}x},
\label{eq_p}
\end{equation}%
with
\begin{equation}
\lambda _{1,2}=\frac{\delta \pm \sqrt{\delta
^{2}+{a_0^2}D\gamma ^{2}}}{D{a_0}}
\end{equation}
{and analogously for $x \in [0, \alpha \l]$, with coefficients
$A',B'$ in (\ref{eq_p}). The unknowns $A,B,A',B'$ are determined
from two conditions (\ref{BCusual}) at steps $x=0,\l$ and
from continuity of current (\ref{surface_current}) and
density $p(x)$ at $x=\alpha \l$.
Before going further, let
us remind what is the relevant function determining the stability
of the evaporating surface. Each step moves with a velocity
proportional to the sum of the upper and lower step current,
$J_{\hbox{\tiny step}}$, as given by Eq.~(\ref{BCusual}).
Perturbing a perfect train of steps, we find~\cite{note_phi} the
stability being governed by the function $\phi (\l )=p(\l )-p(0)$:
the system is stable (unstable), if $\partial _{\l }\phi (\l )$ is
positive (negative). }

We now assume a
diverging drift in the region close to the descending step ($\delta ^{\prime
}\rightarrow \infty $), whose size vanishes in such a way that
the product
\begin{equation}
\frac{\delta ^{\prime }\alpha \l }{{a_0}D}=c  \label{eq_c}
\end{equation}%
is constant.
This model, as discussed at length in the experimental Section, is motivated by a
possible strong impact on drift of a dopant higly localized near steps.
As shown in detail in the Supplemental Material~\cite{supp_mat},
we don't really need a ``singular" model with a diverging $\delta'$
in order to get reversal stability, but the experimental system does
show a singular behavior.

We have also considered the case of a diverging $D$,
but it results to have no effect when restricted to a region of
vanishing size. For this reason, the parameter $D^{\prime }$ does
not appear in the following equations. For ease of notation, {
we define the scaled quantities $\tilde\delta=\delta/(a_0 D)$,
$\tilde\gamma=\gamma/\sqrt{D}$, and $\tilde K=K/D$,  all of
dimension $[L^{-1}]$. }

Solving the two coupled diffusion problems in the above limit, we have found
that the resulting effect is equivalent to introduce a new boundary
condition at the descending step ($x=0$),
\begin{equation}
J_{\hbox{\tiny step}}=-K(p_{\hbox{\tiny step}}e^{-2c}-p_{\hbox{\tiny eq}}) .
\label{BCnewDESCENDING}
\end{equation}
The new condition arises because the density profile in proximity of the
descending step becomes infinitely steep in the limit $\alpha \l %
\rightarrow0 $, changing from some value $p_{\hbox{\tiny step}}$ at $x=0$ to
a value $p_{\hbox{\tiny step}} e^{2c}$ at $x=\alpha \l \rightarrow 0^+$.
Therefore, $p_{\hbox{\tiny step}} e^{2c}$ becomes our new postulated value $%
p(x=0)$, after taking the limit $\alpha \l
\to 0$. { Because of this redefinition of $J_{\hbox{\tiny
step}}$, the function $\phi (\l )$, which governs the stability of
surface dynamics, has to be redefined as $\phi (\l )=p(\l
)-p(0)e^{-2c}$. }

By solving Eq.~(\ref{eq_diffusione}) with usual boundary condition (\ref%
{BCusual}) at $x=\l $ and new boundary condition (\ref%
{BCnewDESCENDING}) at $x=0$, and assuming
\begin{equation}
{
\tilde\delta \l \ll\tilde\gamma \l \ll 1 ,  \label{approximation_linear}
}
\end{equation}
we obtain
\begin{equation}
{
\partial_{\l }\phi(\text{$\l $ })=\frac{2\tK^{2}e^{-2c}p_{\hbox{\tiny eq}%
}(4\tilde\delta-\tK (1-e^{-2c}))}{(\tK +\tK e^{-2c}+2\tilde\delta \tK\l +\tK^{2}e^{-2c}\l )^{2}}.
}
\label{eq_d_phi}
\end{equation}

Therefore, the instability condition $\partial_\l \phi(\l ) <0$ reads
\begin{equation}
{
4\tilde\delta<\tK (1-e^{-2c}),
}
\label{InstabilityCondDescendingStep}
\end{equation}
which reduces to the usual condition $\tilde\delta<0$, when boundary layer
is absent ($c=0$).
It is worthnoting that Eq.~(\ref{InstabilityCondDescendingStep}) does not depend on the
desorption rate $\gamma$. In fact, it is possible to study a model where evaporation is
neglected: in this case, simpler calculations allow to keep $\delta'$ large, but finite.
This calculation, which gives the same result and therefore proves the robustness
of our model, can be found in the Supplemental Material~\cite{supp_mat}.
In the experimental system~\cite{Latyshev} we consider here, desorption is not negligible,
see Eq.~(\ref{approximation_linear}), but we have shown it does not affect the stability reversal process.

If Eq.~(\ref{InstabilityCondDescendingStep}) is satisfied, the instability
time, $\tau_{\hbox{\tiny ins}}$, is set by the relation~\cite{reviewWilliams1999}
\begin{equation}
{
\tau_{\hbox{\tiny ins}}^{-1} = a_0 K |\partial_\l \phi(\l) | .
}
\label{tau_ins}
\end{equation}

Analogously, one may consider a situation with diverging drift in the region
close to the ascending step, $\delta^{\prime}\l (1-\alpha)/{(a_0 D)}=c_{1}$ as $%
\alpha\rightarrow1$. This inhomogeneity results in a new effective boundary
condition at the ascending step,
\begin{equation}
J_{\hbox{\tiny step}}=K(p_{\hbox{\tiny step}}e^{2c_{1}}-p_{\hbox{\tiny eq}})
\end{equation}
and in the redefinition  $\phi(\l )=p(\l )e^{2c_1}-p(0)$, so that
the  instability condition $\partial_\l \phi(\l ) <0$  becomes
\begin{equation}
{
4\tilde\delta<\tK (e^{2c_{1}}-1).
}
\label{InstabilityCondAscendingStep}
\end{equation}

Expressions (\ref{InstabilityCondDescendingStep}) and
(\ref{InstabilityCondAscendingStep}) are a central result of our
paper. In the following we discuss their application  to a
surprising experimental effect, observed on the evaporating
surface of Si(111) in the presence of a tiny coverage on Au atoms.

\textit{The experiment}~\cite{Latyshev}.---Step bunching instability of a
Si(111) vicinal surface heated by a DC electric current to high temperatures
is a well established phenomenon \cite{Latishev89,Metois-Stoyanov}. In the
so-called first temperature regime ($830^\circ$C$\leq T\leq 950^\circ$C),
the regular vicinal surface is stable (unstable) for an uphill (downhill)
current. The effect is usually understood \cite{sequence} by assuming that
neutral silicon adatoms acquire an effective positive charge which gives
rise to an electromigration force $\delta$. In the first temperature regime,
approximations (\ref{approximation_linear}) are valid~\cite%
{approximation_linear}, the density profile is linear, $%
p(x)\simeq\delta x$, and we have stability for $\delta>0$.

A few years ago, it was observed by Kosolobov, Latyshev and
collaborators (KL) that a submonolayer deposition of gold on a
vicinal Si(111) surface drastically affects its stability, which
changes \textit{four} times as a function of the increasing Au
coverage \cite{Latyshev}. {In particular, the addition of a
very small quantity of Au atoms to a clean Si surface (0.0016 ML
in Ref.~\cite{Latyshev98}) resulted in a reversal of stability
(uphill DC setup becomes unstable and viceversa).
Such critical density, not  enough even to
decorate vicinal steps with a
single line of Au atoms, hints at a phase transition at zero critical density, $\rho_{%
\hbox{\tiny Au}}^{\hbox{\tiny cr}}\rightarrow0$ as $\l \rightarrow \infty$.
Moreover, the instability develops much faster
than is usual at this temperature for a clean Si
surface \cite{Latyshev_private}. }

We propose to relate this phenomenon to our model of Fig. 1(b),
with a diverging drift. To do so, {we note that Au adatoms
have a much larger effective charge than Si
adatoms~\cite{Williams97,Yasunaga92}, so that their dynamics in
the electric field is faster. As a consequence, Si adatoms move in
an established environment of Au adatoms, } whose steady profile
may be strongly inhomogeneous near the border. Relating Au
inhomogeneity to Si diffusion inhomogeneity causes a stability
reversal, as demonstrated below.

The key observation is that the negative effective charge of Au atoms and
their strong affinity~\cite{affinity} to the step region, make them
concentrate near the descending step boundary $x=0$. The effective
equilibrium profile of Au adatoms, $n(x)$, is governed by an equation of
type (\ref{eq_diffusione}), i.e.
\begin{equation}
D_{\hbox{\tiny Au}} n^{\prime\prime}(x)+2 {a_0^{-1}}|\delta_{\hbox{\tiny Au}}|
n^{\prime}(x)-\frac{n}{\tau_{\hbox{\tiny Au}}} =0.  \label{AuDiffusion}
\end{equation}

Due to strong affinity assumption, the boundary condition is
$ n(0)=n_{\hbox{\tiny Au}}^{+}$,
where $n_{\hbox{\tiny Au}}^{+}$ is the equilibrium density of Au
adatoms close to the descending step. {Within the
experimental instability time scale, Au coverage remains
approximately constant. Therefore, we can solve
(\ref{AuDiffusion}) assuming a negligible Au evaporation and get }
$n(x)=n_{\hbox{\tiny Au}}^{+}e^{-\kappa x}$,
where $\kappa =2{a_0^{-1}}|\delta _{\hbox{\tiny Au}}|/D_{\hbox{\tiny Au}}$.

If we now impose that the space integral of $n(x)$ equals the total number of
deposited Au atoms,
\begin{equation}
\rho _{\hbox{\tiny Au}}\l =\int_{0}^{\l }n(x)dx=\frac{n_{\hbox{\tiny Au}}^{+}%
}{\kappa }, \label{MassBalance for AscendingStep}
\end{equation}%
we get the relation $\kappa =n_{\hbox{\tiny Au}}^{+}/\rho_{\hbox{\tiny Au}}\l $,
which diverges for $\rho_{\hbox{\tiny Au}}\to 0$.
Finally, the equilibrium density profile writes
\begin{equation}
n(x)=n_{\hbox{\tiny Au}}^{+}\exp \left( -\frac{n_{\hbox{\tiny Au}}^{+}x}{%
\rho _{\hbox{\tiny Au}}\l }\right) .  \label{n(x) for DescendingStep}
\end{equation}

\begin{figure}[ptb]
\includegraphics[width=\columnwidth,clip=yes]{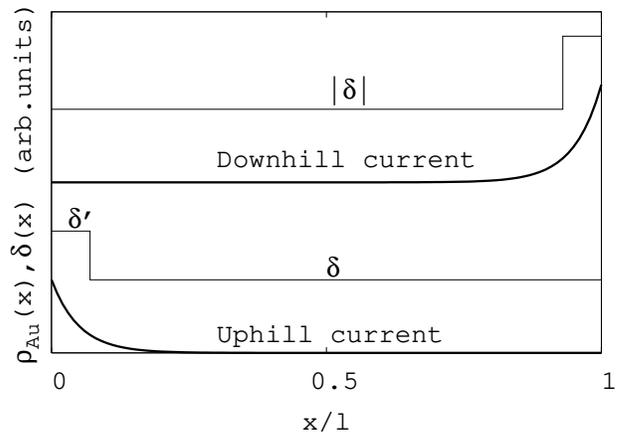}
\caption{Schematic profiles of the density of dopant adatoms
(thick lines) and drifting forces $\protect\delta(x)$ (thin lines)
at small dopant concentration, $\protect\rho_{\hbox{\tiny Au}} \l
\ll 1$. For uphill (downhill) current $\protect\delta>0$(
$\protect\delta<0$), the dopants are concentrated near the left
(right) boundary.  The $\protect\delta<0$ curves are shifted for
better vision. } \label{Fig_AuProfiles}
\end{figure}

Analogously, for downhill current and small density of Au adatoms we obtain
a distribution function peaked near the upper step boundary, $x=\l $,
\begin{equation}
n(x)=n_{\hbox{\tiny Au}}^{-}\exp \left( \frac{n_{\hbox{\tiny Au}}^{-}(x-l)}{%
\rho _{\hbox{\tiny Au}}\l }\right) .  \label{n(x) for AscendingStep}
\end{equation}

As anticipated, we relate the gradient in the density of Au atoms
to an additional drift (effective electromigration force) of Si
adatoms as ~\cite{Krug91}
\begin{equation}
\delta ^{\prime }=\delta -c^{\ast }\frac{\partial n(x)}{\partial x}.
\label{delta_prime(x)}
\end{equation}%
This renormalizes the drift in the region where the gradient is sufficiently
large (see Fig.~2), getting
\begin{equation}
\delta (x)=
{\delta'=c^* n_{\hbox{\tiny Au}}^{+}\kappa}  \text{,~~ for }x\in \lbrack 0,\kappa ^{-1}].
\label{delta(x)}
\end{equation}
If we compare Eq.~(\ref{delta(x)}) with Eq.~(\ref{eq_c}) and identify
the size $(\alpha\l)$ of the vanishing region of divergent drift with
$\kappa^{-1}$, we find the relation $c= c^* (n_{\hbox{\tiny Au}}^{+}/a_0 D)$.
In conclusion, we have related the problem of Si diffusion in the
presence of a tiny quantity of Au doping to the model of
inhomogeneous diffusion, discussed in the previous part of this
Letter.

{We stress that our results
do not  change qualitatively, if the parameter $c$ has a slow
dependence on other quantities such as the local gradient or
average density of Au adatoms: the decisive role is played by the
sign of $c$. Note also that the limit of a diverging drift and
gradient is a simplifying assumption chosen for clarity of
presentation. Simulation of a discrete model with strictly {\it
finite} drifts and gradients yields qualitatively the same results
as the continuous model (to be discussed elsewhere). }

\textit{Discussion}.---The parameters given in Refs.~\cite%
{YangFuWilliams96,LiuWeeks98} lead to dimensionless estimates
{ $\tilde\delta \l \approx 10^{-6},\tilde\gamma \l \approx
10^{-4}$ and $\tK\l \approx 10^{-2}$, }
where $\l \approx 100
$ nm is a typical size of a terrace. This means that assumptions (\ref%
{approximation_linear}) are satisfied and
Eqs.~(\ref{eq_d_phi}-\ref{InstabilityCondAscendingStep}) do apply
to the KL experiments.

For a tiny quantity of gold, Eqs.~(\ref{InstabilityCondDescendingStep}) and (%
\ref{tau_ins}) should be evaluated. Since ${\tK/\tilde\delta } \approx 10^{4}$, Eq.~(%
\ref{InstabilityCondDescendingStep}) is surely satisfied for
finite $c$ (more precisely, for $c>2\times 10^{-4}$). The
microscopic origin of the strong Au-induced increasing of the
effective drift $\delta (x)$ for silicon adatoms may be
 a hard core exclusion
interaction between Au and Si adatoms: the presence of Au adatoms
near the step excludes the presence there of Si adatoms and thus
generates an effective enhanced drift of Si adatoms close to the
step~\cite{Krug91}, proportional to the local density gradient, as
in (\ref{delta_prime(x)}).

Another microscopic origin might be the recharging
effect~\cite{Latyshev98}, according to which a neutral Au adatom
subtracts negative charge from the substrate, while the substrate,
to compensate, transfers equal charge of the opposite sign to the
Si adatoms. Since $q_{\hbox{\tiny Si}}^{\hbox{\tiny eff}}\approx 0.004e$~%
\cite{Williams97} and $q_{\hbox{\tiny Au}}^{\hbox{\tiny eff}}\approx -e$~%
\cite{Yasunaga92}, we expect that recharging effect may lead to a
strong renormalization of $\delta $. Note that the drift
enhancement due to recharging is
 proportional to the local density of the Au adatoms itself, and not to its gradient as
in case of the hard-core exclusion effect alone. With some
amendments to our line of argument (\ref{MassBalance for
AscendingStep})-(\ref{delta(x)}), the instability reversal can be
obtained. We note that both recharging and hard core exclusion
effects contribute with the same sign to the renormalization
(\ref{delta_prime(x)}), leading to the discussed above kinetic
instability reversal.  We also stress that the effective charge of
Si adatoms remains positive across all the transitions.

A piece of evidence that $c$ is not small and also that our theory
does apply to KL experiments is the expectation
that Au doping leads to a much stronger instability than simple
reversing of the sign of the current. In fact, since ${\tK
(1-e^{-2c})\gg \tilde\delta} $, according to
Eqs.~(\ref{eq_d_phi},\ref{tau_ins}) we have that $\tau
_{\hbox{\tiny ins}}^{\hbox{\tiny doping}}\ll \tau _{\hbox{\tiny
ins}}^{\hbox{\tiny reversal}}$, as it is actually seen in
experiments~\cite{Latyshev_private}.

In the Introduction we have stressed that the same doping induces a
change of stability for both signs of the current.
Therefore, let's now consider a downhill current, which means
$\delta<0$ and that the clean Si surface is bunching-unstable, see Eq.~(\ref%
{InstabilityCondAscendingStep}) with $c_1=0$.
Au adatoms are now driven
towards the ascending step and the boundary layer (inhomogeneity) forms at
the ascending step, with a profile given by Eq.~(\ref{n(x) for AscendingStep}%
) (see also Fig.~\ref{Fig_AuProfiles}). Assuming an additional drift of
Si adatoms, $\delta^{\prime}=-c^*\frac {\partial n(x)}{\partial x}%
{= c^* n_{\hbox{\tiny Au}}^{-} \kappa}$,
in the region of a strong Au gradient, $(\l -x)<1/\kappa$,
we recover our model with diverging drift $\delta^{\prime}$ at the ascending
step.
The same
considerations we did for positive $\delta$ and
Eq.~(\ref{InstabilityCondDescendingStep}) are
now applicable to negative $\delta$ and Eq.~(\ref%
{InstabilityCondAscendingStep}), providing perfectly symmetrical conclusions
for a negative electromigration force: a tiny amount of Au doping allows to
stabilize the surface and the stability is much stronger than the stability
gained by simply reversing the field.

Let's now pass to reason on further features of KL experiments~\cite%
{Latyshev}, which go beyond our simple model. These experiments are done for a
variable quantity of Au, from a clean Si surface ($\rho _{\hbox{\tiny Au}}=0$%
) to an almost full covering ($\rho _{\hbox{\tiny Au}}=1$)~\cite{note_100}
and authors find a total of four stability transitions, while increasing $%
\rho _{\hbox{\tiny Au}}$ (see Fig.~3). It is reasonable to assume that
enhanced drift is suppressed with increasing Au coverage, which becomes more
homogeneous. Since a model with a constant $\delta (x)$ is equivalent to a
clean Si surface, we expect that increasing $\rho _{\hbox{\tiny Au}}$ leads
to a new reversal of stability, in agreement with experiments. However, why
a further increase of $\rho _{\hbox{\tiny Au}}$ produces a further change of
stability? This change of stability is better understood if large $\rho_{%
\hbox{\tiny Au}}$ coverages are described in terms of \textquotedblleft Au
holes", i.e. empty places which can potentially host Au adatoms: $\rho _{%
\hbox{\tiny Au}}^{\hbox{\tiny holes}}=1-\rho _{\hbox{\tiny Au}}$. A gradient
of Au adatoms corresponds to a gradient of Au holes, which appear to have a
strong \textit{positive} effective charge. This change of sign compensates
the minus relating the two spatial derivatives,
$\rho_{\hbox{\tiny Au}%
}^{\prime\hbox{\tiny holes}}= -\rho_{\hbox{\tiny Au}}^{\prime}$.
Thus, removing a tiny quantity of Au from
a fully covered Si surface reverses its stability.
In Fig.~3 we graphically summarize the stability diagram
of Au-doped Si(111).

\begin{figure}[tbp]
\includegraphics[width=\columnwidth,clip=yes]{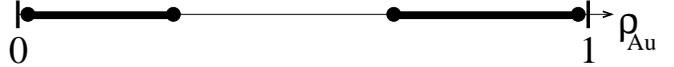}
\caption{Stability diagram of a Si(111) surface, with an uphill current ($%
\protect\delta>0$), when doped with a variable quantity of Au. The clean
surface ($\protect\rho_{\hbox{\tiny Au}}=0$) and the fully covered Si
surface ($\protect\rho_{\hbox{\tiny Au}}=1$) are stable. A tiny Au
deposition on the clean surface or a tiny removal of it from the fully
covered surface induces a stability reversal and the surface becomes
unstable (thick horizontal lines). In the intermediate region, the surface
is expected to recover stability (thin horizontal line), because the Au
profile is more flat. If the sign of the current is reversed, stability
regions are also reversed (thick line $\Leftrightarrow$ thin line). }
\end{figure}

We conclude with one prediction of our model: if dopant
adatoms have a positive effective charge, such as Cu
or Ag \cite{Williams97}, the stability reversal effect is not
expected. Indeed, positively charged foreign adatoms under uphill
current ($\delta>0$) will be driven to a ascending step and
eventually form a profile (\ref{n(x) for AscendingStep}), for
which (\ref{InstabilityCondAscendingStep}) will  apply. Proceeding
along the lines (\ref{delta_prime(x)}-\ref{delta(x)}), we conclude
that (\ref{InstabilityCondAscendingStep}) cannot be satisfied,
since $\delta>0$ and $c_1<0$, so that an instability reversal will
not happen. Indeed, for a
Cu-doped surface, the reversal of bunching stability was not observed \cite%
{Latyshev98}.

\textit{Acknowledgements}.---VP acknowledges financial support from the
italian MIUR (Ministero dell'Istruzione, dell'Universit\`{a} e della
Ricerca) through PRIN 20083C8XFZ initiative.


\begin{thebibliography}{99}
\bibitem{note_100} In the experiment, there is a maximal monolayer density
of Au doping, which depends on several factors (e.g., Au surface
reconstruction~\cite{Au-Si} or miscut~\cite{affinity}) and which is of order
0.75~\cite{Latyshev}. In our model, the maximal covering of Au was
renormalized to $\rho_{\hbox{\tiny Au}}^{\hbox{\tiny max}}=1$.
\end{thebibliography}
\end{document}